\documentclass[12pt,dvips]{article}
\usepackage{bm}
\usepackage{hyperref}
\usepackage{graphicx}
\usepackage{amsmath}
\usepackage{amssymb}
\begin{document}
\begin{flushright}
JLAB-THY-11-1445
\end{flushright}
\begin{center}
{\LARGE Nuclear physics with a medium--energy \\[0ex]
Electron--Ion Collider\footnote{Mini--review compiled in preparation
for the MEIC Conceptual Design Report, Jefferson Lab (2011).}}
\\[3ex]
{\large
A. Accardi$^{1,2}$, V. Guzey$^2$, A. Prokudin$^2$, C. Weiss$^2$} 
\\[2ex]
\small $^1$\ Hampton University, Hampton, Virginia 23668, USA \\[0ex]
\small $^2$\ Jefferson Lab, Newport News, Virginia 23606, USA
\\[2ex]
October 4, 2011
\end{center}
\begin{abstract}
A polarized $ep/eA$ collider (Electron--Ion Collider, or EIC) with variable
center--of--mass energy $\surd s \sim 20-70$ GeV and a luminosity 
$\sim 10^{34} \, \textrm{cm}^{-2} \textrm{s}^{-1}$ would be uniquely
suited to address several outstanding questions of Quantum Chromodynamics
(QCD) and the microscopic structure of hadrons and nuclei: (i)~the 
three--dimensional structure of the nucleon in QCD (sea quark and gluon
spatial distributions, orbital motion, polarization, correlations); 
(ii)~the fundamental color fields in nuclei (nuclear parton densities, 
shadowing, coherence effects, color transparency); (iii)~the conversion of 
color charge to hadrons (fragmentation, parton propagation through matter,
in--medium jets). We briefly review the conceptual aspects of these questions 
and the measurements that would address them, emphasizing the 
qualitatively new information that could be obtained with the collider. 
Such a medium--energy EIC could be realized at Jefferson Lab after the 
12~GeV Upgrade (MEIC), or at Brookhaven National Lab as the low--energy 
stage of eRHIC.
\end{abstract}
{\bf Introduction.}
Understanding the internal structure of hadrons and nuclei on the basis 
of the fundamental theory of strong interactions, Quantum Chromodynamics 
(QCD), is one of the central problems of modern nuclear physics, as 
described in the 2007 NSAC Long--Range Plan~\cite{2007NSAC}. It is the key to 
understanding the dynamical origin of mass in the visible universe and 
the behavior of matter at astrophysical temperatures and densities. 
It is an essential step in describing nuclear structure and reactions
from first principles, with numerous applications to science and technology.
Theoretical methods to apply QCD to hadronic and nuclear systems have 
made dramatic advances in the last two decades but rely crucially
on new experimental information for further progress.

Electron scattering has been established as a powerful tool 
for exploring the structure of matter at the 
sub--femtometer level ($< 1 \, \textrm{fm} = 10^{-15}\,\textrm{m}$). 
Historically, 
such experiments provided the first proof of the extended nature 
of the proton and revealed the presence of pointlike constituents, or quarks, 
at smaller scales, revolutionizing our understanding of strong interactions.
Subsequent experiments established the validity of QCD and the presence
of gluonic degrees of freedom at short distances and measured the basic
number densities of quarks and gluons in the nucleon (proton, neutron). 
While much progress has been made, several key questions remain 
unanswered~\cite{2007NSAC}:
\begin{itemize}
\item[I)] What role do non--valence (``sea'') quarks and gluons play
in nucleon structure? What are their spatial distributions? How do they 
respond to polarization? What is their orbital motion, and how does 
it contribute to the nucleon spin? The answers to these questions will 
provide essential information on the effective degrees of freedom
emerging from QCD at distances of the order of the hadronic size 
($\sim 1\, \textrm{fm}$).
\item[II)] What are the properties of the fundamental QCD color fields 
in a nucleus? What are the nuclear gluon and sea quark densities?
To what extent are they modified by nuclear binding, 
quantum--mechanical interference, and other collective effects?
These questions are the key to understanding the QCD origins
of nucleon interactions at different energies, the role of 
non--nucleonic degrees of freedom, and the approach to a 
new regime of high gluon densities and saturation at high energies.
\item[III)] How do colorless hadrons emerge from the colored quarks 
and gluons of QCD? What dynamics governs color neutralization and
hadron formation? By what mechanisms does the color charge of QCD 
interact with nuclear matter? We are still far from understanding the 
strong interaction dynamics of the fundamental process of conversion 
of energy into matter.
\end{itemize}

It is now widely accepted that a polarized $ep/eA$ collider 
(Electron--Ion Collider, or EIC) with a 
variable $ep$ center--of--mass (CM) energy in the range 
$\surd s = 20-70 \, \textrm{GeV}$, and a luminosity of
$\sim 10^{34} \, \textrm{cm}^{-2} \textrm{s}^{-1}$ over most
of this range, would offer a unique opportunity to address 
these questions~\cite{Boer:2011fh}. Such a facility would provide the 
necessary combination of kinematic reach (resolution scale, energy span), 
luminosity (precision, multi--dimensional binning, rare processes), 
and detection capabilities (resolution, particle identification) 
to study nucleon 
and nuclear structure through scattering experiments with a variety of 
final states. It would represent the natural next step after the 
high--luminosity fixed--target $ep/eA$ experiments 
(JLab 12 GeV, SLAC) and the high--energy HERA $ep$ collider 
(protons only, unpolarized). It would be the first ever high--energy
electron--nucleus collider and open up qualitatively new 
possibilities to study QCD in the nuclear environment. Finally, 
polarized beams would allow one to investigate proton and neutron spin 
structure with unprecedented accuracy and kinematic reach; 
such measurements were so far possible only in
fixed--target experiments (EMC, SMC, SLAC, HERMES, COMPASS, JLab)
or polarized $pp$ collisions (RHIC). In the following we
summarize what measurements with such a medium--energy EIC could 
contribute to answering the above questions.
Nuclear physics at higher energies and possible studies of
electroweak interactions with an EIC are described in Ref.~\cite{Boer:2011fh}.

%
%
\begin{figure}
\centering
\parbox[c]{0.53\textwidth}{\includegraphics[width=0.53\textwidth]
{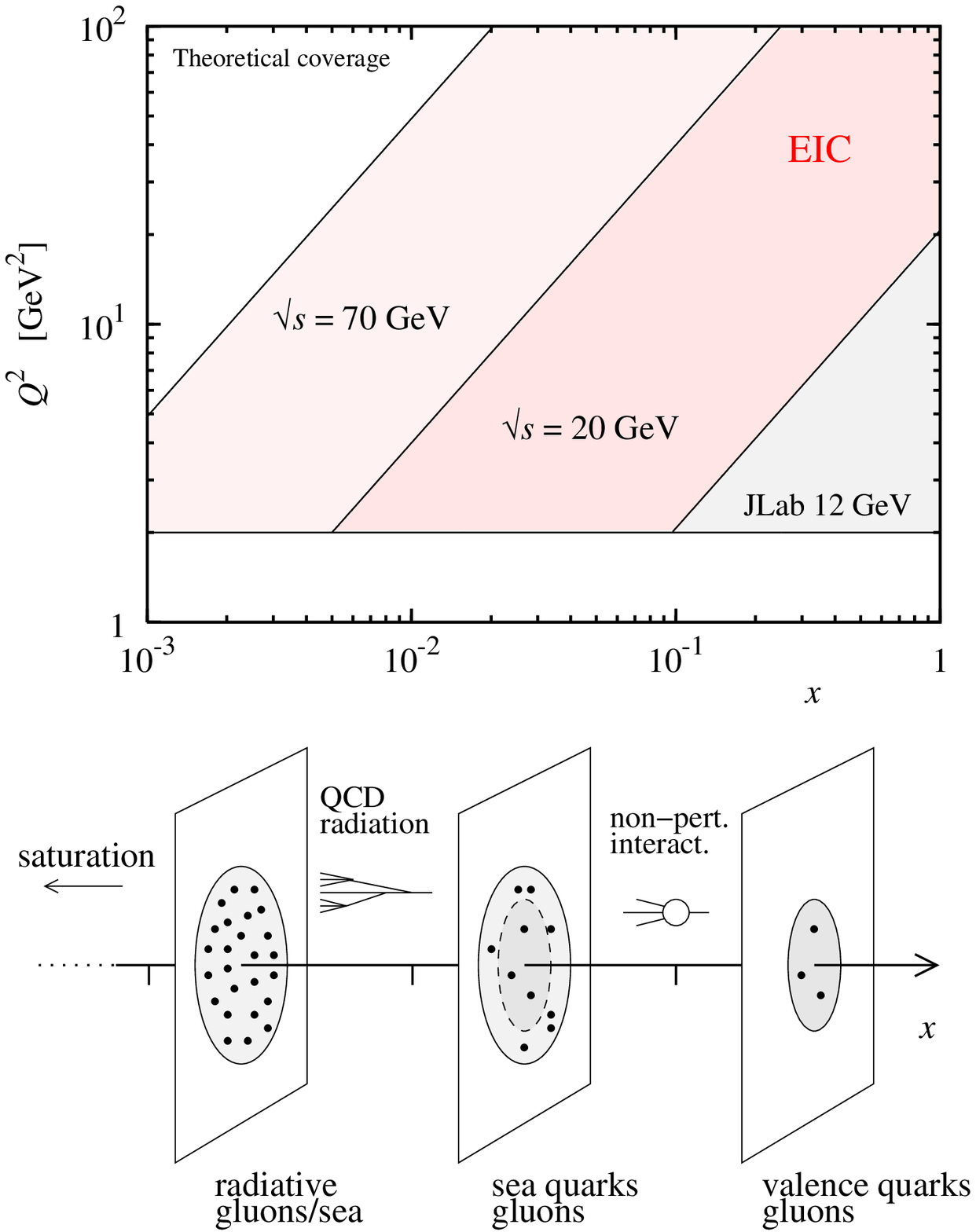}}
\hspace{0.05\textwidth}
\parbox[c]{0.37\textwidth}{
\begin{center}
\includegraphics[width=0.30\textwidth]{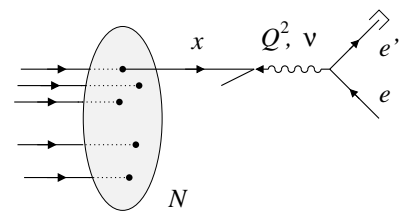}
\end{center}
\caption{\small
\textit{Top right:} Deep--inelastic electron--nucleon scattering.
The momentum transfer $Q^2$ defines the resolution scale, the Bjorken 
variable $x$ is the momentum fraction of the constituents probed
in the scattering process, and $\nu = Q^2/(2Mx)$ determines the 
boost imparted to the struck quark and the hadrons emerging from
its fragmentation.
\textit{Top left:} Kinematic coverage in $x$ and $Q^2$ with 
JLab 12 GeV and a medium--energy EIC ($\surd s =$ 20 and 70 GeV), for 
$Q^2_{\rm min} = 2\, \textrm{GeV}^2$. 
\textit{Bottom left:} Components of the nucleon wave function probed 
in scattering experiments at different $x$ (see axis on graph).}
\label{fig:kinplane}
}
\end{figure}

{\bf Three--dimensional structure of the nucleon in QCD.}
The nucleon in QCD represents a dynamical system of fascinating complexity. 
In the rest frame it may be viewed as an ensemble of interacting color
fields, coupled in an intricate way to the vacuum fluctuations that 
govern the effective dynamics at distances $\sim 1\, \textrm{fm}$. In this 
formulation its properties can be studied through large--scale numerical 
simulations of the field theory on a discretized space--time 
(Lattice QCD) \cite{USQCD}
as well as analytic methods. A complementary description 
emerges when one considers a nucleon that moves fast, with a momentum 
much larger than that of the typical vacuum fluctuations. In this limit 
the nucleon's color fields can be projected on elementary quanta with 
point--particle characteristics (partons), and the nucleon becomes a 
many--body system of quarks and gluons. As such it can be described by a 
wave function, in much the same way as many--body systems in nuclear 
or condensed matter physics (see Fig.~\ref{fig:kinplane}). In contrast 
to these non--relativistic systems, in QCD the number of pointlike 
constituents is not fixed, as they constantly undergo creation/annihilation 
processes mediated by QCD interactions, reflecting the essentially 
relativistic nature of the dynamics. A high--energy scattering process 
takes a ``snapshot'' of this fast--moving system with a spatial resolution 
given by the inverse momentum transfer $1/Q$. The energy transfer, 
parametrized by the Bjorken variable $x$, defines the momentum fraction
of the struck constituent and thus determines what particle 
configurations are intercepted in the scattering process. In this
way one can probe in detail the various components of the
wave function and map out their properties (see Fig.~\ref{fig:kinplane}).
Measurements with JLab 12 GeV probe nucleon structure in the
region dominated by the valence quark component ($x > 0.1$),
including the unknown $x \rightarrow 1$ region~\cite{CDR}. 

In addition to the valence quarks, the nucleon contains 
a ``sea'' of quark--antiquark pairs that is created by non--perturbative 
QCD interactions and reflects the complex structure of the ground state 
(or vacuum) of the theory. The spin and flavor quantum numbers carried 
by the sea sit mainly in the region $0.01 \lesssim x < 0.1$ and are 
poorly constrained by present data. A medium--energy EIC could measure 
the distribution of sea quarks through semi--inclusive measurements, 
in which the charge and flavor of the struck quark/antiquark are ``tagged'' 
by detecting hadrons ($\pi^\pm, K^\pm, p, \bar p, \ldots)$ produced from 
its fragmentation. Compared to fixed--target experiments, the 
energy available with the collider ensures that the hadronization of the 
struck quark proceeds independently from the target remnants and cleanly 
preserves the original spin--flavor information. The kinematic coverage 
and detection capabilities are uniquely suited to such measurements, 
allowing for a precise mapping of this largely unexplored component 
of the nucleon.

Equally important is the distribution of polarized gluons in the nucleon.
Besides its intrinsic importance, its measurement is needed 
to solve the ``puzzle'' of the nucleon spin decomposition and 
quantify the role of orbital angular momentum in the nucleon 
wave function. Since gluons carry no electric charge, electromagnetic
scattering can probe them only indirectly, through the $Q^2$
dependence of the nucleon structure functions. Present $eN$ data,
together with those from polarized $pp$ collisions at RHIC, 
practically do not constrain the polarized gluon density for 
$x \lesssim 0.05$. Inclusive measurements with a medium--energy EIC 
would dramatically extend the data set and determine the polarized nucleon 
structure function $g_1(x,Q^2)$ down to $x \sim \textrm{few} \times 10^{-3}$ 
with a substantial range in $Q^2$ (see Fig.~\ref{fig:kinplane}), 
allowing one to extract the polarized gluon density from the 
$Q^2$ dependence \cite{Burkardt:2008jw}.

%
%
\begin{figure}
  \centering
  \parbox{0.40\textwidth}{\includegraphics[width=0.4\textwidth]{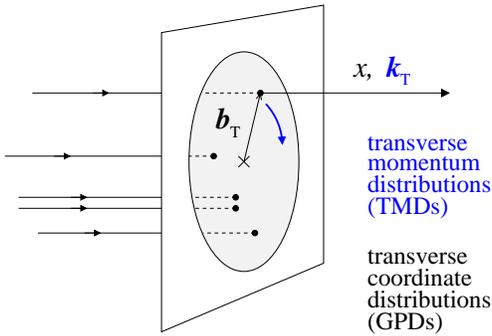}}
  \hspace{0.04\textwidth}
  \parbox{0.54\textwidth}{\caption{\small
Three--dimensional structure of the fast--moving 
nucleon in QCD. The distribution of partons (quarks, gluons) is
characterized by the longitudinal momentum fraction $x$ and the
transverse spatial coordinate $\bm{b}_T$ (GPDs). In addition, the 
partons are distributed over transverse momenta $\bm{k}_T$, 
reflecting their orbital motion and interactions in the system (TMDs). 
Polarization distorts both the spatial and momentum distributions. 
Note that $\bm{b}_T$ and $\bm{k}_T$ are not 
Fourier conjugate; a joint description in both variables can be 
formulated in terms of a Wigner phase space density. 
Observables sensitive to either $\bm{b}_T$ or $\bm{k}_T$ help to 
establish a three--dimensional dynamical picture of the nucleon in QCD.}
\label{fig:TMD}
}
\end{figure}
Other fundamental characteristics of the nucleon are the transverse
spatial distributions of quarks and gluons carrying a certain momentum 
fraction $x$ (see Fig.~\ref{fig:TMD}).
They define the basic size and ``shape'' of the nucleon in QCD and 
convert the one--dimensional picture conveyed by the longitudinal momentum 
densities into a full three--dimensional image of the fast--moving 
nucleon~\cite{GPD_reviews}. 
Information on the transverse distribution of quarks and gluons
is obtained from exclusive scattering $\gamma^\ast N \rightarrow M + N
\, (M = \textrm{meson}, \gamma, \textrm{heavy quarkonium})$. Such 
processes probe the generalized parton distributions (GPDs), which 
combine the concept of the quark/gluon momentum density with that of
elastic nucleon form factors. Measurements of $J/\psi$ photo/electroproduction
with a medium--energy EIC would be able to map the transverse spatial 
distribution of gluons in the nucleon above 
$x \sim \textrm{few} \times 10^{-3}$ in unprecedented detail.
In particular, these measurements would cover the unexplored gluons 
in the valence region at $x \gtrsim 0.1$, whose presence has been inferred
from global fits to deep--inelastic scattering data but has rarely been
confirmed directly; their dynamical origins are one of the outstanding
questions of nucleon structure in QCD. Information on the transverse
spatial distribution of gluons is needed also to describe the final states 
in $pp$ collisions at LHC (underlying event in hard processes, multiparton 
processes) and understand the approach to the regime of high gluon densities 
at small $x$ (initial conditions for non--linear QCD evolution 
equations) \cite{Frankfurt:2005mc}. Measurements of real photon 
production ($\gamma$, deeply virtual Compton scattering) with an EIC would 
differentiate gluon and quark spatial distributions and study how the 
latter are deformed in a transversely polarized nucleon. 
Production of light mesons with charge/isospin ($\pi, K, \rho,
K^\ast$) would map the transverse distributions of sea quarks and
provide additional insight into their dynamical origins. This program
of ``quark/gluon imaging'' requires differential measurements of
low--rate processes and relies crucially on the high luminosity
provided by the EIC in the envisaged energy range,
and the possibility to longitudinally and transversely
polarize the proton beam.

Closely related is the question of the orbital motion of quarks and
gluons and its role in nucleon structure (see Fig.~\ref{fig:TMD}). 
This information is encoded in the transverse momentum distributions 
(TMDs) and their response to nucleon and quark/gluon 
polarization~\cite{Anselmino:2011ay}. They provide a three--dimensional 
representation of the nucleon in momentum space, complementing the 
spatial view offered by the GPDs. The TMDs can be measured in 
semi--inclusive scattering processes $\gamma^\ast N \rightarrow h + X$, 
where particles produced by fragmentation of the struck quark 
($h = \pi, K$, $J/\psi$, open charm), as well as the nucleon fragments, 
can reveal the quark and gluon transverse momentum and its correlation 
with the nucleon spin. The various structure functions, each of which
describes certain facets of nucleon structure (transverse motion and 
deformation, spin--orbit correlations, orbital angular momentum, 
final--state interactions of the struck quark with the color fields
in the nucleon) can be separated by measurements with different 
combinations of beam and target polarizations, including transverse 
nucleon polarization easily available with the collider. 
Measurements with a medium--energy 
EIC will be able to precisely determine, \textit{e.g.}, the valence and sea 
quark Sivers function sensitive to spin--orbit interactions 
in the region $x > 0.01$, where it is expected to be sizable. 
They will also study for the first time the $Q^2$ evolution of TMDs 
and the region of large transverse momenta, $k_T \gg 1\, \textrm{GeV}$, 
where TMDs can be related to multiparton correlations in the nucleon.
Measurements with open charm and $J/\psi$ mesons in the final state 
can directly probe the gluon TMDs. All these studies require 
multi--dimensional binning in $x$, $Q^2$, and the energy
fraction $z$ and transverse momentum $P_T$ of the produced meson,
which can be performed only with high--statistics data as would become 
available with the planned EIC luminosity.

\begin{figure}[t]
\parbox[c]{0.55\textwidth}
{\includegraphics[width=0.53\textwidth]{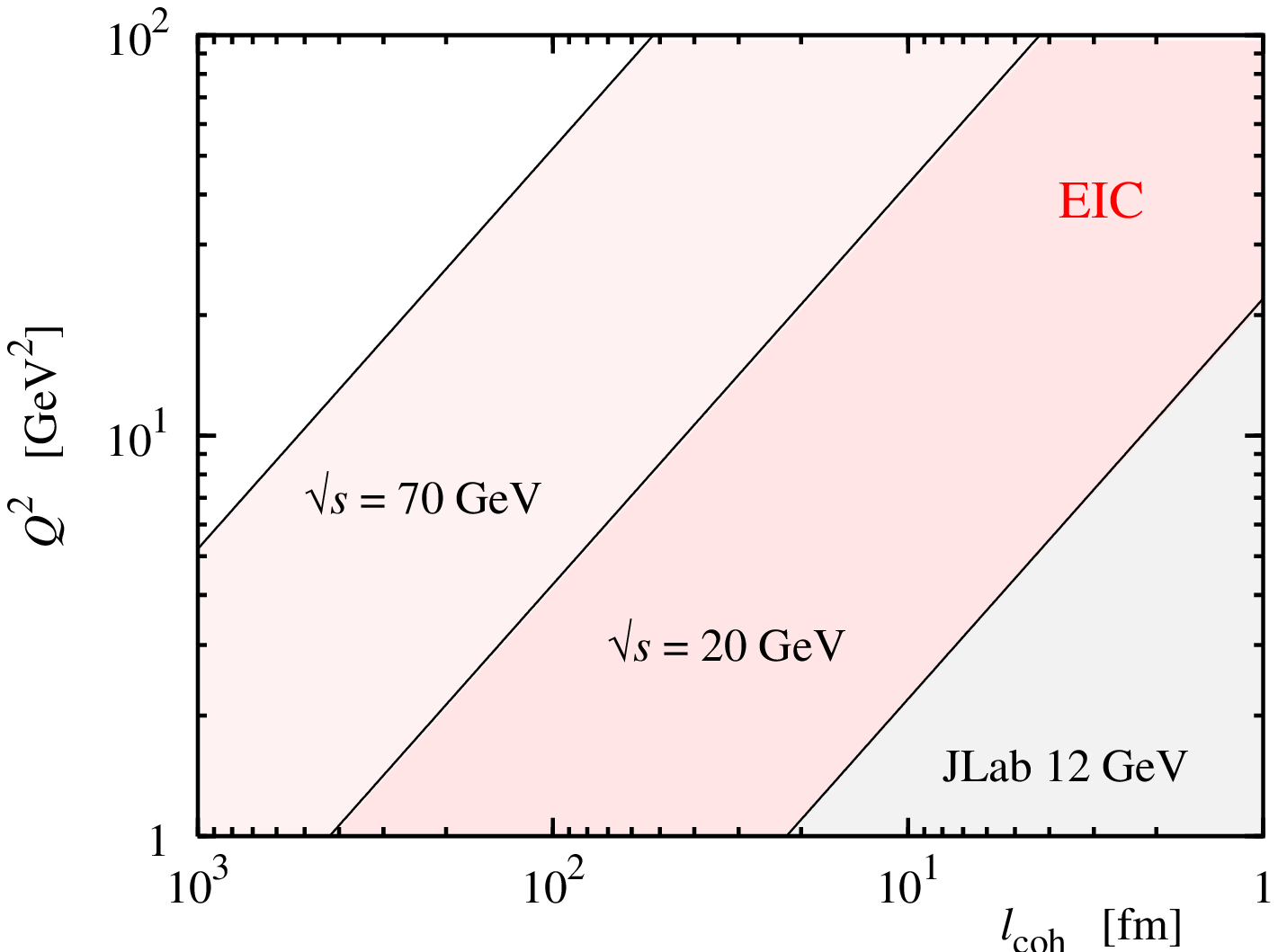}
\\[1ex]
\includegraphics[width=0.51\textwidth]{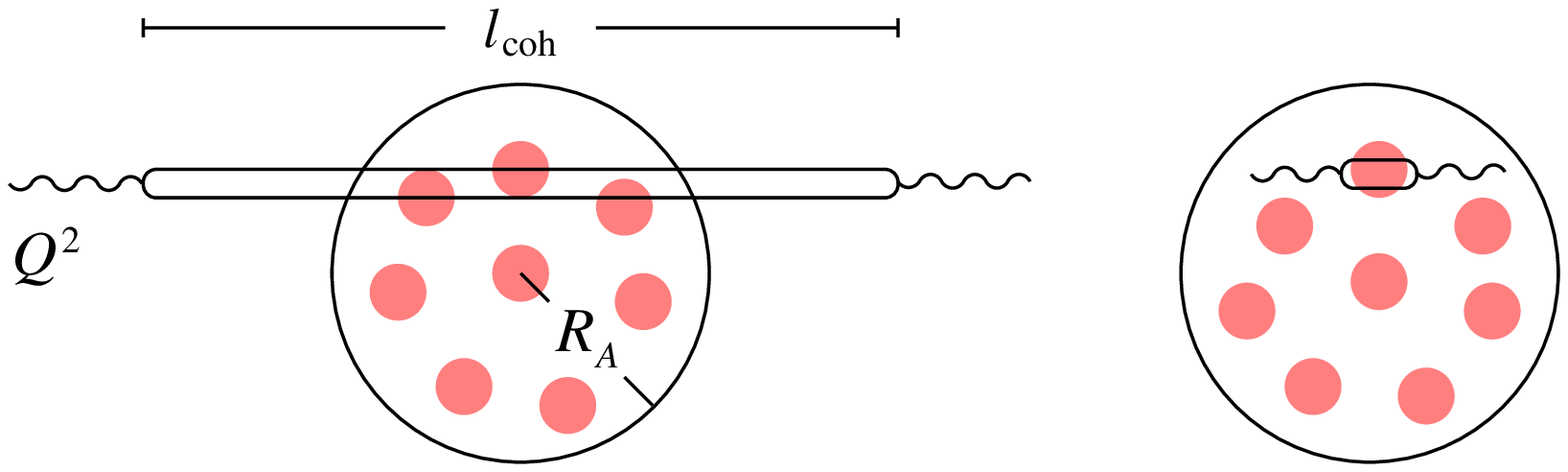}}
\hspace{0.03\textwidth}
\parbox[c]{0.37\textwidth}{
\caption{\small
\textit{Top:} Kinematic coverage in the typical coherence length 
$l_{\rm coh} = (M_N x)^{-1}$ and
the scale $Q^2$ in electron--nucleus scattering with a medium--energy
EIC. \textit{Bottom:} Interaction of a small--size probe with a nucleus.
The coherence length defines the lifetime of the probe in the rest frame
of the nucleus, \textit{i.e.}, the longitudinal extent of the
interaction region. If $l_{coh} \ll R_A$ the interaction is mainly with 
the color field of a single nucleon, possibly modified by nuclear 
binding (\textit{bottom right}). If $l_{coh} \gtrsim R_A$ the probe 
can interact coherently with the color fields of all nucleons located 
at the transverse position of impact (\textit{bottom left}).}
\label{fig:coherence}
}
\end{figure}

{\bf Color fields in nuclei.} A basic quest of nuclear physics is to 
understand the structure and dynamics of the QCD color fields in nuclei 
with nucleon number $A > 1$. Information on these fields is obtained
by studying the scattering of small--size probes --- \textit{e.g.}, 
a virtual photon with $Q^2 \gg 1\, \textrm{GeV}^2$ --- from nuclei 
over a range of incident energies (see Fig.~\ref{fig:coherence}).
Of particular interest is how the nuclear fields differ from the
sum of the color fields of the individual nucleons. The lifetime of the 
probe in the target rest frame is defined by the coherence length, 
$l_{\rm coh} \propto (x M_N)^{-1}$, where the coefficient depends on the 
size of the probe but is typically of order unity. If the coherence 
length is much smaller than the nuclear radius $R_A \sim \textrm{few fm}$, 
the scattering involves only a single nucleon in the nucleus. In this 
regime the results can be interpreted in terms of a modification of 
single--nucleon structure through nuclear binding and reveal the QCD origins 
of the nucleon--nucleon interaction. If the coherence length becomes 
comparable to or larger than the nuclear radius, $l_{\rm coh} \gtrsim R_A$, 
the color field seen by the probe is the quantum--mechanical superposition 
of the fields of the individual nucleons, resulting in a rich spectrum 
of coherence effects such as shadowing \cite{Frankfurt:2011cs}, 
diffraction, and eventually the approach to the unitarity limit 
(saturation) at high energies \cite{Gelis:2010nm}. In addition, the fields 
change with the energy and the resolution scale $Q^2$ as a result of QCD 
radiation. An EIC with a CM energy in the range 
$\surd s \sim 20-70\, \textrm{GeV}$ would for the first time provide 
the coverage in $l_{\rm coh}$ and $Q^2$ necessary to observe these 
phenomena, opening up a whole new area of study.
%
%

Much interesting information already comes from the basic quark and 
gluon densities of nuclei. A peculiar pattern of nuclear 
modifications was observed in fixed--target experiments and caused 
much excitement; it shows suppression compared to the free nucleon
for $0.2 < x < 0.8$ (the famous ``EMC  effect,'' to be explored
further with JLab 12 GeV), some signs of 
enhancement for $0.05 < x < 0.2$, and significant 
suppression at smaller $x$ (``shadowing'') \cite{Norton:2003cb}. 
However, such experiments were unable to reach deep in the shadowing 
region, distinguish valence and sea quarks, or probe gluons.
Nuclear deep--inelastic scattering with a medium--energy EIC would 
for the first time allow one to determine the gluon and sea quark 
densities in a range of nuclei. Thanks to the wide kinematic 
coverage, the EIC will be able to penetrate deep into the shadowing
region, while simultaneously having sufficient $Q^2$ range  
to extract the nuclear gluon densities through the $Q^2$ dependence
of the structure function $F_2^A$ at both small and large $x$. 
Measurements at different energies would isolate 
the longitudinal structure function $F_{L}^A$, which provides
direct access to gluons. Using a combination of 
inclusive measurements and gluon tagging through charm production,
an EIC will be able to explore nuclear gluons also in the antishadowing 
and EMC effect regions --- a step that might prove as revolutionary for 
our understanding of nuclei as the discovery of the quark EMC effect 30 
years ago. 

Further information on the nuclear modification of the quark/gluon 
structure of the proton and the neutron can be gained from 
deep--inelastic measurements with detection of the spectator system
of $A - 1$ nucleons in the final state. In particular, scattering
from the deuteron with a tagged spectator proton can measure the 
structure functions of the bound neutron at controlled virtualities, 
from which the free neutron quantities can be obtained by extrapolation 
to the on--shell point. Measurements with a tagged spectator neutron, 
which are extremely difficult with a fixed target but feasible with 
a collider using a zero degree calorimeter, provide completely new 
information on the bound proton structure functions that constrains 
theoretical models of binding effects and the on--shell extrapolation
(contrary to the neutron, the free proton structure function is known
from independent measurements with a proton target).
Tagged measurements on heavier nuclei could explore the effects 
of nucleon embedding in a complex nuclear environment. Measurements
of coherent nuclear scattering, in which the nucleus remains intact and
is detected with a small recoil momentum of $\lesssim 100\, \textrm{MeV}$
in the final state, can map the transverse gluonic radius of nuclei 
and study shadowing as a function of the impact parameter --- information
essential for the analysis of high--energy $pA$ and $AA$ collisions.

Experiments with nuclear targets also provide qualitatively new insight 
into the short--distance dynamics of deep--inelastic processes.
A fundamental prediction of QCD as a gauge theory is color transparency:
the interaction of small--size colored configurations with hadronic
matter is governed by their color dipole moment and vanishes proportionally 
to their transverse size. A medium--energy EIC would allow one to test
this prediction through measurements of meson electroproduction on nuclei
over a wide range of $l_{\rm coh}$ and $Q^2$, controlling the 
longitudinal extent of the interaction region and the transverse 
size of the $q\bar q$ configuration. Previous fixed--target measurements
(E665, HERMES) could not vary these parameters fully independently.

Detailed studies of color transparency and a reliable determination of 
the nuclear gluon density in the shadowing region $0.001 < x < 0.1$,
as envisaged with a medium--energy EIC, are essential also 
for a quantitative assessment of the approach to the saturation 
regime at small $x$. In this regime the transverse density of gluons
interacting with a high--energy probe becomes so large that it constitutes 
a new dynamical scale that can serve as the basis for systematic 
calculations of inclusive cross sections as well as final--state
characteristics \cite{Gelis:2010nm}. Saturation dynamics is expected to 
be important in $AA$ and central $pp$ collisions at the LHC and has
been associated with phenomena observed in heavy--ion collisions at RHIC. 
Nuclear shadowing, as would be established with a medium--energy EIC, may 
slow down the approach to gluon saturation at small $x$ 
\cite{Frankfurt:2011cs}. The study of the saturation regime proper will be 
the object of high--energy colliders such as a high--energy EIC ($eA$)
or the LHeC ($ep$ and $eA$, see below). This program would involve 
measurements of inclusive and diffractive nuclear structure functions at 
small $x$, analysis of final states in which the saturation scale could 
manifest itself directly (\textit{e.g.}, $p_T$ spectra of leading 
forward particles), and measurements of multiparticle correlations
sensitive to the dynamics of the radiation processes generating the 
dense gluon medium.

\begin{figure}
\centering
\includegraphics[width=0.6\textwidth] {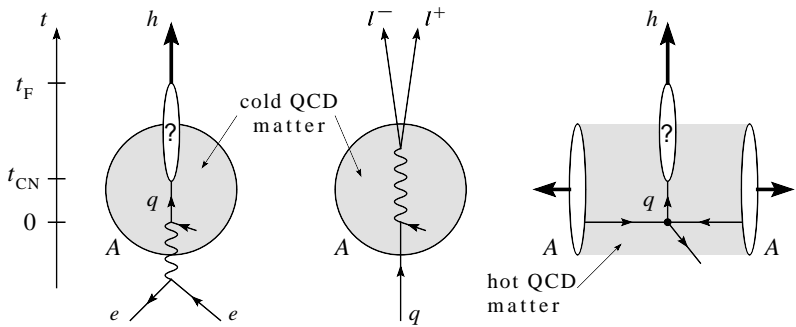}
\caption{\small
Parton propagation and hadronization in cold and hot nuclear matter. 
The color neutralization ($t_{\rm CN}$) and hadron formation ($t_{\rm F}$) 
time scales are indicated on the vertical time axis.
\label{fig:coldhot}
}
\end{figure}

{\bf Emergence of hadrons from color charge.}
The emergence of colorless hadrons from the elementary color charge
produced by short--distance probes --- the so--called hadronization process 
--- is a principal aspect of QCD which still lacks a quantitative 
understanding from first principles \cite{Accardi:2009qv}.
Empirical fragmentation functions, which encode the probability
that a quark or gluon decays into a hadron and colored remnant, 
have been obtained by fitting experimental data, but knowledge of the 
underlying dynamics remains sketchy and model--dependent. Basic questions 
concern even the characteristic time scales for the neutralization of 
color charge (sometimes referred to as pre--hadron formation) and the 
formation of physical hadrons (see Fig.~\ref{fig:coldhot}). 
Measuring these time scales would be the first step toward understanding 
how hadrons emerge dynamically from the color charge of QCD, 
complementing the information obtained from hadron structure and spectroscopy
studies.

Nuclear deep--inelastic scattering provides a known and stable
nuclear medium (``cold QCD matter'') and a final
state with good experimental control of the kinematics of the hard
scattering. This permits one to use nuclei as femtometer--scale
detectors of the hadronization process (see Fig.~\ref{fig:coldhot}).  
By measuring the energy loss and transverse momentum broadening
of leading hadrons induced by the nuclear environment one can
discriminate the different dynamical processes (medium--induced gluon 
bremsstrahlung, pre--hadron reinteraction) and infer the space--time evolution 
of hadronization. Theoretical models of these processes can be calibrated 
in $eA$ scattering and then applied to study, \textit{e.g.,}\ the 
Quark--Gluon Plasma created in high--energy nucleus--nucleus collisions
(``hot QCD matter,'' see Fig.~\ref{fig:coldhot}). 
The combination of high energy and luminosity offered by the EIC
promises a truly qualitative advance in this field, compared with
current and planned fixed target experiments. 
The large $Q^2$ range permits measurements in the fully calculable
perturbative regime with enough leverage 
to determine nuclear modifications in the QCD evolution of
fragmentation functions; the high luminosity permits 
the multidimensional binning necessary for separating the
many competing effects and for detecting rare
hadrons. The large energy range $\nu \approx 10-1000\,\textrm{GeV}$ 
allows one to experimentally boost the hadronization process in 
and out of the nuclear medium, in order to cleanly extract the
color neutralization and hadron formation times (small $\nu$)
and isolate in--medium parton propagation effects (large $\nu$). 
The quark and gluon  in--medium energy loss measured in this way is of 
major interest in its own right, as it addresses the fundamental question 
of the interaction of an energetic color charge with hadronic matter
in QCD. With an EIC one will be able for the first time to study also the 
in--medium propagation and hadronization of heavy quarks (charm, bottom)
in $eA$ collisions, which is necessary to test predictions for 
their energy loss and confront puzzling measurements of heavy flavor 
suppression in the Quark--Gluon Plasma at RHIC. 

Furthermore, an EIC with $\surd s \gtrsim 30$ GeV will permit for the first
time to measure jets and their substructure in $eA$ collisions.
The modifications compared to jets in $ep$ scattering in the same kinematics
can be related to the propagation of the colored parton shower 
in the nuclear medium and offers new insight into its space--time
evolution. It can also be used to measure the cold nuclear matter 
transport coefficients which encode basic information on the 
non--perturbative gluon fields in nuclei. 

Another interesting aspect of hadronization is the evolution of the 
system from which a color charge has been removed by the 
hard process. In deep--inelastic $ep$ scattering with an EIC at
$\surd s \sim 20-70 \, \textrm{GeV}$ one would for the first time
be able to cleanly separate the virtual photon (or current) and 
target fragmentation regions in the final state and study the properties 
of the latter using forward detectors. In this way one could follow the 
materialization of the ``color hole'' in the nucleon created by the
hard process. Measurements of particle correlations between the current
and target fragmentation regions (\textit{e.g.}, particles originating from 
$s$ and $\bar s$ quarks) would provide new insight into the nucleon's 
spin--flavor structure and could reveal dynamical pair correlations
in the nucleon's partonic wave function, as are expected to be induced
by the dynamical breaking of chiral symmetry in the QCD vacuum.

\begin{figure}
\parbox[c]{0.59\textwidth}{
\includegraphics[width=0.58\textwidth]{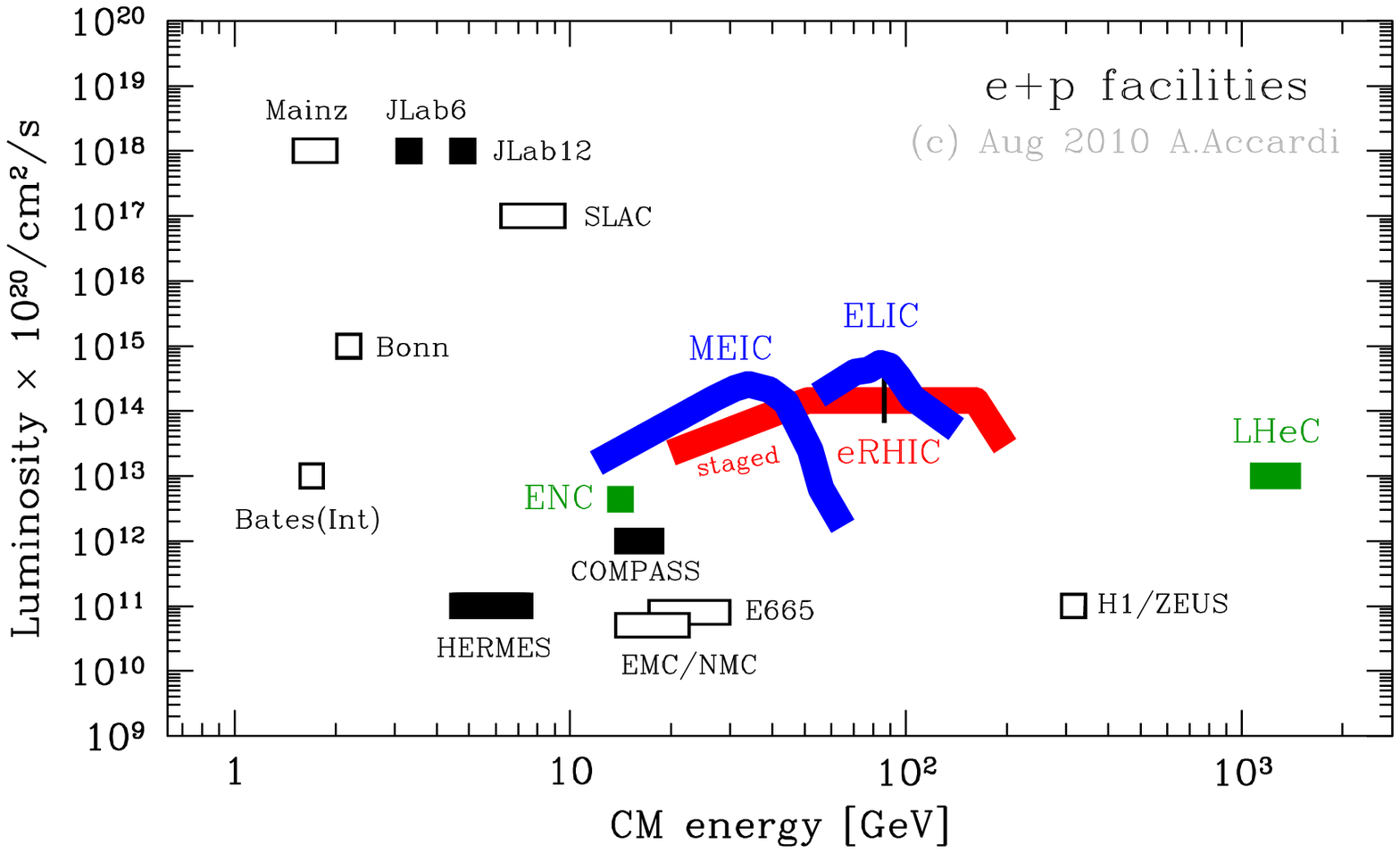}
 \\[-1ex]
\includegraphics[width=0.58\textwidth]{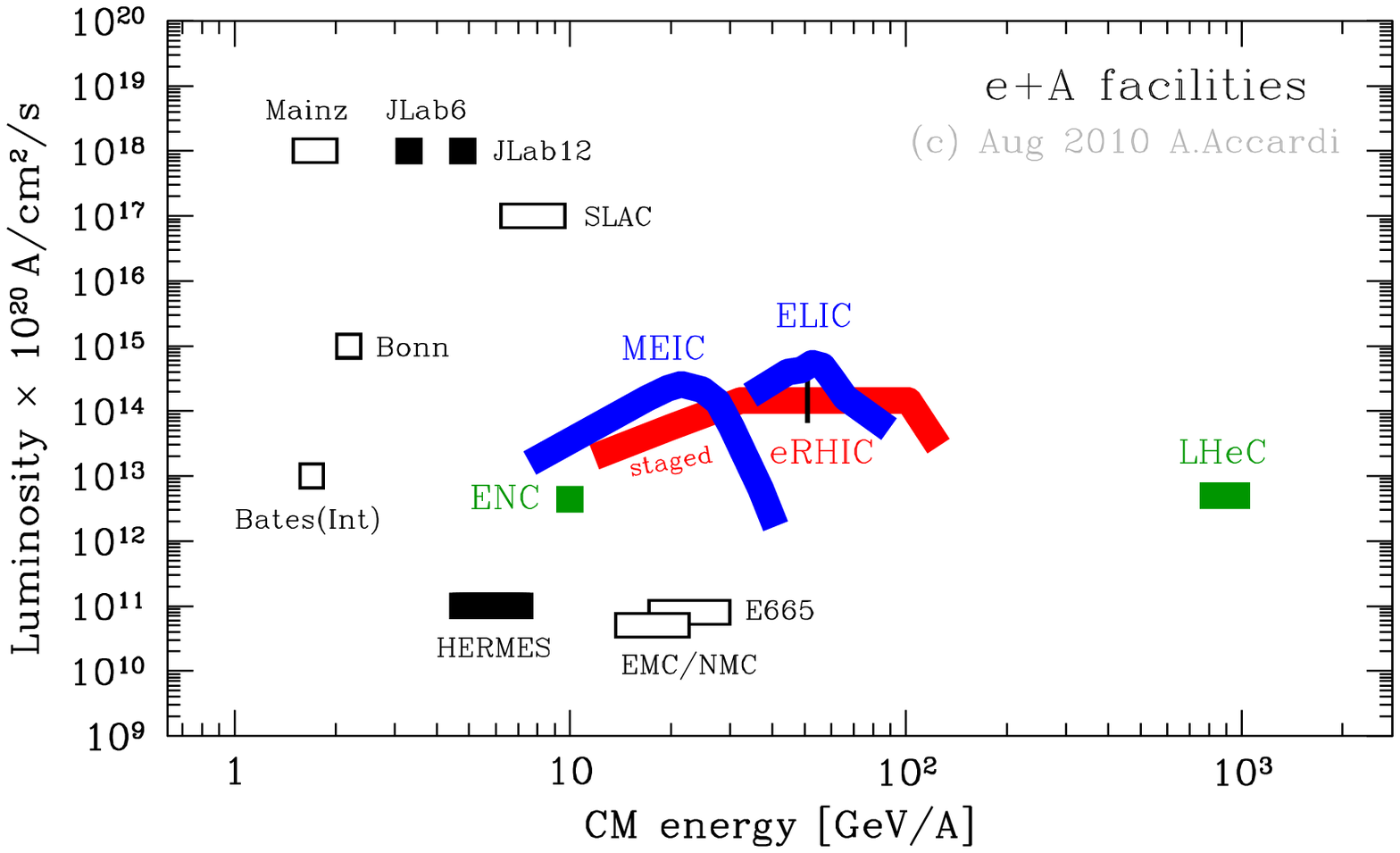}
}
\hspace{0.03\textwidth}
\parbox[c]{0.32\textwidth}{
\caption{\small
Projected luminosity for $ep$ (top) and $eA$ (bottom) collisions as a 
function of the CM energy per nucleon for the JLab (MEIC, ELIC; blue)
and BNL (eRHIC; red) designs as of August 2010 \cite{EIC-lumi,EIC-designs}. 
Also shown are the values achieved by existing and past $ep/eA$ facilities,
as well as the projections for the planned $ep/eA$ collider at CERN 
(LHeC) \cite{LHeC} and the low--energy $ep$ collider at GSI (ENC) 
\cite{ENC}.
\label{fig:lumi}
}}
\end{figure}
{\bf Possible realizations of a medium--energy EIC.}
Two scenarios for the realization of a medium--energy EIC of 
$\surd s \sim 20 - 70 \, \textrm{GeV}$ are presently being discussed.
The design proposed by Jefferson Lab (MEIC) would use the 11 GeV CEBAF 
electron accelerator and a newly built ion complex as injectors for 
a ring--ring $ep/eA$ collider with energies $E_e = 3-11\, \textrm{GeV}$ 
and $E_p = 20-100\, \textrm{GeV}$ and a circumference of 
$\sim 1\, \textrm{km}$, slightly smaller than that of the present 
CEBAF accelerator. This design would achieve a luminosity of
the order of $10^{34}\, \textrm{cm}^{-2} \, \textrm{s}^{-1}$ over a
broad range of CM energies. The ring is laid out in the form 
of a Figure--8 for optimal polarization transport and could 
support up to four interaction points. A high--energy collider of 
$\surd s \sim 70-100 \, \textrm{GeV}$ (ELIC) could be realized as an
upgrade with a larger ring of $\sim 2.5\, \textrm{km}$.
The design proposed by Brookhaven National Lab (eRHIC) would 
use the proton/ion beam of RHIC with energy up to 325 GeV and collide 
it with a 5~GeV electron beam accelerated by energy--recovering linacs placed
along a new recirculating ring in the RHIC tunnel. Higher CM energies
could be realized by increasing the electron energy from 5 to 20 
(possibly 30) GeV through addition of  superconducting radio--frequency 
cavities to the linacs (``staging''). Figure~\ref{fig:lumi} shows the
projected luminosity for $ep$ and $eA$ collisions as a function of the 
CM energy per nucleon for both designs (the figures shown here reflect 
the status of the designs as of August 2010 and were the basis for the
physics simulations compiled in Ref.~\cite{Boer:2011fh}). Comparison with 
the values achieved with existing or past $ep/eA$ facilities shows that a 
medium--energy EIC would dramatically extend the combined 
``energy $\times$ luminosity'' frontier, enabling the next--generation
nuclear physics experiments described in this summary. Details of the 
accelerator designs and further information can be found 
in Ref.~\cite{EIC-lumi,EIC-designs}.

{\bf Acknowledgments.} We thank R.~Ent, P.~Nadel--Turonski, and the 
JLab EIC study group for the suggestion to prepare this mini--review 
and several informative discussions. 
The work of A.~A.\ is partially supported by NSF Award No.~1002644.
Notice: Authored by Jefferson Science Associates, LLC under U.S.\ DOE
Contract No.~DE-AC05-06OR23177. The U.S.\ Government retains a
non--exclusive, paid--up, irrevocable, world--wide license to publish 
or reproduce this manuscript for U.S.\ Government purposes.
\end{document}